\documentclass[%
 pra,
 amsmath,amssymb,
 reprint,%
]{revtex4-1}

\usepackage{graphicx}% Include figure files
\usepackage{dcolumn}% Align table columns on decimal point
\usepackage{bm}% bold math
\begin{document}
%\title{Improved faked-state attack on Entanglement-Based Quantum Key Distribution}
\title{Double blinding-attack on entanglement-based quantum key distribution protocols}

\author{Guillaume Adenier}
\email{adenier@rs.noda.tus.ac.jp}
\author{Masanori Ohya}%
\author{Noboru Watanabe}%
\affiliation{Tokyo University of Science, 2641 Yamazaki, Noda, Chiba 278-8510, Japan}%
\author{Irina Basieva}
\author{Andrei Yu. Khrennikov}
\affiliation{Linnaeus University, Vejdes plats 7, SE-351 95 V\"{a}xj\"{o}, Sweden}%

\begin{abstract}
We propose a double blinding-attack on entangled-based quantum key distribution protocols. The principle of the attack is the same as in existing blinding attack except that instead of blinding the detectors on one side only, Eve is blinding the detectors of both Alice and Bob. In the BBM92 protocol, the attack allows Eve to get a full knowledge of the key and remain undetected even if Alice and Bob are using 100$\%$ efficient detectors. The attack can be easily extended to Ekert protocol, with an efficiency as high as 85.3$\%$.
\end{abstract}

\maketitle
Practical implementation of Quantum Key Distribution (QKD) protocols \cite{Gisin02} can be subjected to attacks exploiting the imperfections of the components used by the two parties (Alice and Bob) who wish to generate a shared key to encrypt their communication on a public channel. Most notable attacks are the time-shift attacks \cite{Lo09a,Lo09b} and the blinding-attacks \cite{Makarov07,Makarov10a,Makarov10}. The latter have demonstrated a full hacking of a QKD protocol, the eavesdropper (Eve) acquiring the exact knowledge of the sift key shared by Alice and Bob in a BBM92 protocol \cite{BBM92}.

We propose here an improvement on the existing blinding-attacks by attacking both sides (Alice and Bob) instead of one. The advantage of our proposed attack is that it reaches 100$\%$ efficiency on both sides in the case of BBM92 protocol, and that it can easily be extended to cover the case of Ekert protocol \cite{Ekert91}.

\section{Single blinding-attack}
The existing blinding attacks \cite{Makarov07,Makarov10a,Makarov10} on BBM92 protocol are intercept-and-resend type of attacks. Eve intercepts the signal intended for Bob, and performs measurements in random bases to obtain the raw key, as Bob would have done it.

In order to hide her presence, for each successful measurement result that Eve obtains, she forwards to Bob a signal that deterministically gives him the exact same result whenever their measurement basis are the same, and no result at all (non detection) if they are diagonal to each others.

To implement this idea with actual QKD devices, Eve blinds Bob's detectors to single-photon detection. She does so with various techniques \cite{Makarov10a} by forcing the detectors to exit the Geiger mode  and enter the linear mode, in which the detector clicks only when the intensity of the signal reaching the detectors exceeds the preset discriminator threshold $I_\mathrm{th}$. After each detection, Eve forwards a bright pulse linearly polarized along the direction corresponding to her own measurement result. When the bases chosen at random by Eve and Bob are identical, the pulse deterministically produces a click in one of Bob's detectors. Because it is then either entirely reflected or entirely transmitted at Bob's polarizing beamsplitter, Bob's measurement results are then the same as Eve's. In order to avoid double counts and incorrect results whenever the bases chosen at random by Eve and Bob are diagonal to each others, Eve sets the intensity of the pulses such that it is lower than twice the threshold intensity in the detectors. The pulse is then split in half at Bob's polarizing beamsplitter whenever the bases are diagonal to each other, so that the output in either of Bob's detector is insufficient to overcome the threshold to produce a click.

The point of the attack is that at the end of the raw key distribution, Eve owns an exact copy of Bob's key. So, if Alice and Bob are satisfied enough with the quantum bit error rate (QBER) measured on a subset of this key, Eve can listen to the error correction protocol that they implement and perform the exact same operations as Bob, and can obtain in the end an exact copy of the sifted key \cite{Makarov10}.

A weakness of single blinding-attacks is that Bob's key is on average half the size that he would have normally obtained in the absence of an attack, because in about half of the cases the bases chosen at random by Eve and Bob turn out to be diagonal to each others, and Bob's detectors do not click: the efficiency of this attack is by design fundamentally limited to 50$\%$ on Bob's side.

Another weakness is that extending this type of attack to cover the case of Ekert protocol is not straightforward. To the best of our knowledge, actual attacks against Ekert protocol have yet to be implemented with real devices, and proposals to do so are not entirely satisfactory. For instance, in a proposal using blinding attack \cite{Makarov07}, rates of coincidences that would be expected equal from a genuine source of entangled state differ significantly, and that is something that Alice and Bob would not fail to notice.

The double blinding-attack that we propose here addresses these two weaknesses. The idea is simply to launch a blinding attack on both sides and to drive detection patterns inspired by existing local realist models.

\section{Double blinding-attack}

Regardless of the protocol used by Alice and Bob (BBM92 or Ekert), the implementation of a double blinding-attack on each side is similar to that of a single blinding attack, except that Eve is blinding all detectors instead of only those on Bob's side. Any practical implementation of a QKD protocol that is vulnerable to blinding-attacks \cite{Makarov07,Makarov10a,Makarov10} would thus be immediately vulnerable as well to our proposed attack.

A crucial difference however in the spirit of the attack is that it is not an intercept-and-resend attack: Eve is not measuring anything from the genuinely entangled source that was initially intended for Alice and Bob, let alone using any information that she could possibly extract from this source. Eve is blocking instead this entangled source altogether and replacing it entirely by tailored pairs of bright pulses.

To be more specific, Eve is sending pairs of bright pulses, with intensity $I_\mathrm{A}$ and polarization $\lambda_\mathrm{A}$ for Alice, and intensity $I_\mathrm{B}$ and polarization $\lambda_\mathrm{B}$ for Bob, with the condition $$\lambda_\mathrm{A}=\lambda_\mathrm{B}-\frac{\pi}{2}=\lambda,$$ which guarantees that the measurement results will be correlated.

Eve is randomizing the polarization $\lambda$ from one pulse to the other, using a uniform distribution on the circle in order to obtain an attack that is rotationally invariant, both at the single count level and at the coincidence count level.

Consider Alice's side. By Malus law, the intensity of a pulse linearly polarized along $\lambda$ reaching the detectors 0 and 1 at the output of the polarizing beam-splitter (PBS) oriented along $\theta_\mathrm{A}$ is:

\begin{equation}\label{outputI}
\left\{
\begin{aligned}
    I_\mathrm{A,0}&=I_\mathrm{A} \cos^2(\lambda-\theta_\mathrm{A})=I_\mathrm{A}\frac{1+\cos 2(\lambda-\theta_\mathrm{A})}{2}
\\
    I_\mathrm{A,1}&=I_\mathrm{A} \sin^2(\lambda-\theta_\mathrm{A})=I_\mathrm{A}\frac{1-\cos 2(\lambda-\theta_\mathrm{A})}{2}
\end{aligned}
\right.
\end{equation}

Once it is forced to exit the Geiger mode, a detector clicks in the linear mode only if the signal intensity reaching this detector is greater than the threshold $I_\mathrm{th}$ that was set for the Geiger mode \cite{Makarov10a}. For simplicity we assume that the threshold is the same in all detectors. A click is triggered in detector $i$ if the intensity $I_\mathrm{A,i}$ is such that
\begin{equation}\label{thr}
    I_\mathrm{A,i}>I_\mathrm{th},
\end{equation}
and similarly on Bob's side the condition to obtain a click in detector $i$ is
\begin{equation}\label{thrb}
    I_\mathrm{B,i}>I_\mathrm{th}.
\end{equation}

As we will see, for a given threshold $I_\mathrm{th}$, the only parameter that Eve needs to adjust is the brightness of the pulses, depending on which protocol Alice and Bob are implementing.

\subsection{Attack on BBM92 protocol}\label{BBM92}
In the BBM92 protocol \cite{BBM92}, the security of the key is supposed to be guaranteed by a low enough QBER. The idea is that any measurement performed by Eve meant to extract some information from a source of genuinely entangled photons would necessarily introduce errors in the perfect (anti)correlation predicted for the singlet state.

In the double blinding-attack, Eve is bypassing this difficulty by creating a source from scratch in which she has a full knowledge of the polarization and intensity of the pulses.

 The idea of the attack is to adjust the bright pulses such that the detection pattern behaves exactly like John Bell's local hidden-variable model \cite{Bell64}, which was meant to reproduce the perfect correlation predicted for identical measurement directions on a singlet state \cite{EPR35}.

 Eve does so by adjusting the intensity of her bright pulses such that they have twice the threshold intensity $I_\mathrm{th}$ to obtain a click in a detector:
 $$I_\mathrm{A}=I_\mathrm{B}=2\;I_\mathrm{th}.$$

Then, on Alice's side, the condition (\ref{thr}) to obtain a click with the outputs (\ref{outputI}) becomes:
\begin{equation}
\left\{
\begin{aligned}
        \cos 2(\lambda-\theta_\mathrm{A})&>0 \qquad \textrm{for a click in channel 0,}
\\
        -\cos 2(\lambda-\theta_\mathrm{A})&>0 \qquad \textrm{for a click in channel 1.}
    \end{aligned}
\right.
\end{equation}

Counting a click in channel 0 as a $+1$ and a click in channel $1$ as a $-1$, the measurement result $\mathrm{A}$ for Alice takes the form
\begin{equation}
       \mathrm{A}(\theta_\mathrm{A},\lambda)=\mathrm{sign}\;\cos 2(\lambda-\theta_\mathrm{A}).
\end{equation}

Similarly, for the same pulse, the $\pi/2$ shift in polarization on Bob's side leads to a measurement result of the form
\begin{equation}\label{Bob1}
       \mathrm{B}(\theta_\mathrm{B},\lambda)=-\mathrm{sign}\;\cos 2(\lambda-\theta_\mathrm{B}).
\end{equation}
For a uniform distribution of $\lambda$ over the circle, it leads to a correlation of the form
\begin{equation}
       E(\theta_\mathrm{B},\theta_\mathrm{A})=-1+\frac{4}{\pi}\;|\theta_\mathrm{B}-\theta_\mathrm{A}|,
\end{equation}
where $|\theta_\mathrm{B}-\theta_\mathrm{A}|\in[-\frac{\pi}{2},+\frac{\pi}{2}]$.

The detection pattern of this attack is nothing but that of the local hidden-variable model given by John Bell in his 1964 article \cite{Bell64}, except that the angles given here are half of those given by Bell because he was considering the singlet state for spin 1/2 particles when we are considering photons. A representation in Poincar\'{e} sphere would have given us exactly the same angle dependence.

The important property of this attack for the BBM92 protocol is that whenever Alice and Bob are performing the same measurements $\theta=\theta_\mathrm{A}=\theta_\mathrm{B}$, the correlation is
\begin{equation}
    E(\theta,\theta)=-1,
\end{equation}
which means that their results are perfectly anticorrelated, exactly as predicted for the singlet state for identical measurements. This perfect correlation for identical measurement is all that is needed to achieve a low QBER in a BBM92 protocol.

It is worth noticing here that the conditions to obtain a click in a detector are mutually exclusive, so that there are no double-counts. Even more crucial is that there is always at least one detector that clicks, except in the case of $\lambda=\theta_\mathrm{A}$, which can be ignored since it is a null set, so that the detection efficiency on each side is in effect equal to 100$\%$.

Blinding attacks against the BBM92 protocol are therefore unrelated to the detection loophole, contrary to what was usually thought. The effectiveness of the double blinding-attack rather highlights the intrinsic weakness of the BBM92 protocol which is only probing the perfect correlation in identical bases, something that is always accessible to local realist models (with Bertlmann's socks type models).

\subsection{Attack on Ekert protocol}\label{Ekert}
In Ekert protocol, the security of the key is guaranteed by measuring a high enough violation of Bell inequalities \cite{CHSH,Ekert91}. Extending the attack against BBM92 protocol to cover the case of Ekert protocol is straightforward. All Eve needs to do is lower the intensity of the pulses sent on one side (say, on Alice's side), such that $I_\mathrm{th}<I_\mathrm{A}<2I_\mathrm{th}$. On the other side (Bob's side), the pulses are the same as in the attack on the BBM92 protocol, that is $I_\mathrm{B}=2I_\mathrm{th}$, so that the detection pattern remains as in Eq.~(\ref{Bob1}).

The simple consequence of these less bright pulses on Alice's side is that they do not always generate a click in one of Alice's detectors. This is enough to let her enter the realm of the detection loophole, and leads to an apparent violation of Bell inequalities on the sample of detected pulses (see \cite{Adenier09} and reference therein for an account on the importance of the discriminator threshold in the context of Bell inequalities violations). The lower the intensity $I_\mathrm{A}$ with respect to the fixed threshold $I_\mathrm{th}$, the more pulses fail to trigger a pulse and the higher the violation of Bell inequalities measured on the sample of detected pulses \cite{Adenier09}.

Setting the intensity of the pulses on Alice's side such that
$$I_\mathrm{A}\cos^2\alpha=\;I_\mathrm{th},$$
the condition (\ref{thr}) to obtain a click on Alice's side becomes:
\begin{equation}
\left\{
\begin{aligned}
        \cos 2(\lambda-\theta_\mathrm{A})&>\cos 2\alpha \qquad \textrm{for a click in channel 0,}
\\
        -\cos 2(\lambda-\theta_\mathrm{A})&>\cos 2\alpha \qquad \textrm{for a click in channel 1.}
    \end{aligned}
\right.
\end{equation}
so that whenever the condition
\begin{equation}
    -\cos 2\alpha<\cos 2(\lambda-\theta_\mathrm{A})<\cos 2\alpha
\end{equation}
is fulfilled, neither detector clicks.

Counting a click in channel 0 as a $+1$, a click in channel $1$ as a $-1$, and a non-detection as $0$, the measurement result $\mathrm{A}$ for Alice becomes:
\begin{equation}\label{weakerpattern}
\left\{
\begin{aligned}
    \mathrm{A}(\theta_\mathrm{A},\lambda)&=0 \qquad \textrm{ when } \alpha<|\lambda-\theta_\mathrm{A}|<\frac{\pi}{2}-\alpha,
\\
    \mathrm{A}(\theta_\mathrm{A},\lambda)&=-\mathrm{sign}\;\cos 2(\lambda-\theta_\mathrm{A}) \qquad \textrm{otherwise},
    \end{aligned}
\right.
\end{equation}
for $(\lambda-\theta_\mathrm{A})\in [-\pi/2,\pi/2]$, and the correlation measured by Alice and Bob on the sample of detected pulses then takes the form
\begin{equation}\label{corr}
\left\{
\begin{aligned}
    E(\theta_\mathrm{B},\theta_\mathrm{A})&=-1 \qquad \textrm{for } 0\leq|\theta_\mathrm{B}-\theta_\mathrm{A}|<\frac{\pi}{4}-\alpha,
\\
    E(\theta_\mathrm{B},\theta_\mathrm{A})&=+1 \qquad \textrm{for } \frac{\pi}{4}+\alpha<|\theta_\mathrm{B}-\theta_\mathrm{A}|<\frac{\pi}{2},
\\
    E(\theta_\mathrm{B},\theta_\mathrm{A})&=\frac{1}{\alpha}\;\big(|\theta_\mathrm{B}-\theta_\mathrm{A}|-\frac{\pi}{4}\big) \qquad \textrm{otherwise,}
    \end{aligned}
\right.
\end{equation}
for $(\theta_\mathrm{B}-\theta_\mathrm{A})\in [-\pi/2,\pi/2].$

With $\alpha=\frac{\pi}{4\sqrt{2}}$, the above correlation leads to a correlation with magnitude $\frac{\sqrt{2}}{2}$ for the angle differences $|\theta_\mathrm{B}-\theta_\mathrm{A}|=\frac{\pi}{8}+k\frac{\pi}{4}$ used in Ekert protocol, and that means a violation of the Bell-CHSH inequalities measured on the sample of detected pulses of $S_\mathrm{CHSH}=2\sqrt{2}$, which is the maximum predicted by Quantum Mechanics for a singlet state \cite{CHSH}.

It is worth noticing that this attack designed for Ekert protocol works without change if Alice and Bob perform a BBM92 protocol instead, because the correlation given by Eqs.~(\ref{corr}) when they perform identical measurement is equal to $-1$.

Note also that, thanks to the rotational invariance of the source, the marginal probabilities are equal and independent of the settings, and the correlation function depends only on the angle difference between the measurement performed by Alice and Bob, and not on their absolute values. The only way to spot this attack by looking at the statistics of the detected events would therefore consists in monitoring the detection efficiencies.

On the side receiving the weaker pulses (here Alice's side) it is straightforward, using Eqs.~(\ref{weakerpattern}), to evaluate the probability of detection $p_\mathrm{w}$ for a uniform distribution of the polarization $\lambda$:
\begin{equation}\label{eff}
    p_\mathrm{w}=1-\frac{2}{\pi}\int_{\alpha}^{\frac{\pi}{2}-\alpha}d\lambda=\frac{4\alpha}{\pi},
\end{equation}
that is, $p_\mathrm{w}=\frac{1}{\sqrt{2}}$ with $\alpha=\frac{\pi}{4\sqrt{2}}$.

On the side receiving the stronger pulses (here on Bob's side) the probability of detection is equal to 1, so that this imbalance could reveal Eve's attack if left as such. She can however hide this behavior by sending alternatively (or randomly) the weaker pulse on either Alice's side or Bob's side. Then the detection efficiency $\eta$, which is the probability for a pulse to be detected on either output channel, becomes the average of the probability to detect a strong or a weak pulse, that is,
\begin{equation}\label{trueeta}
    \eta=\frac{1}{2}(1+p_\mathrm{w})=\frac{1}{4}(\sqrt{2}+2)\approx0.853.
\end{equation}
 This efficiency is greater than the efficiency bound of $82.8\%$ above which no local realist model reaching the maximum violation of Bell-CHSH inequalities predicted by Quantum Mechanics exists \cite{GargMermin,Larsson98}.

 In order to explain this seemingly surprising behavior, a first point to notice is that the non-detections are not independent in this attack: there is always at most one pulse that remains undetected, the strong one being always detected. So, the proof of the bound given by Garg and Mermin \cite{GargMermin} simply does not apply here.

 A second important point is how the efficiency is actually defined and estimated. It is known that assuming the independence of non-detections is not a necessary condition to derive the bound \cite{Larsson98}, but it is then established for a \emph{conditional} efficiency $\eta_{2,1}$: the probability for a photon to be detected on one side given that its corresponding photon (from the same pair) was detected on the other side. Its relevant Bell-CHSH inequality then has the same form as derived by Garg and Mermin \cite{GargMermin}:
\begin{equation}\label{coineffLarsson}
    S_\mathrm{CHSH}\leq\frac{4}{\eta_{2,1}}-2,
\end{equation}
with $\eta_{2,1}\geq2/3$ \cite{Larsson98}. The maximum value $2\sqrt{2}$ predicted by Quantum Mechanics for $S_\mathrm{CHSH}$ \cite{CHSH} exceeds the right-hand side of this inequality if $\eta_{2,1}>2(\sqrt{2}-1)\approx0.828$. It means a bound on this conditional efficiency of $82.8\%$ to invalidate local realism.

Now, assigning the same bound not just to the conditional efficiency $\eta_{2,1}$ but rather to the actual detection efficiency $\eta$ (which is what we have calculated above for the double-blinding attack) is not immediate.

One can assume independent non-detection events, as was done by Garg and Mermin \cite{GargMermin}, so that $\eta=\eta_{2,1}$ and the bound derived for the conditional efficiency $\eta_{2,1}$ is then valid just the same for the detection efficiency $\eta$.

In a case however in which the assumption of independence is not fulfilled, as with the double-blinding attack, the bound on the conditional efficiency $\eta_{2,1}$ cannot be extended directly to the detection efficiency $\eta$. One needs to use another inequality given by Larsson \cite{Larsson98} between the conditional efficiency and the detection efficiency, that is $\eta_{2,1}\geq2-\frac{1}{\eta}$, which leads to a less stringent Bell-CHSH inequality, this time as a function of the detection efficiency $\eta$:
\begin{equation}\label{deteffLarsson}
    S_\mathrm{CHSH}\leq\frac{2}{2\eta-1},
\end{equation}
with $\eta\geq3/4$ \cite{Larsson98}. The maximum value $2\sqrt{2}$ predicted by Quantum Mechanics for $S_\mathrm{CHSH}$ exceeds the right-hand side of this inequality if $\eta>\frac{1}{4}(\sqrt{2}+2)\approx0.853$. It means a bound on the efficiency $\eta$ of $85.3\%$ to invalidate local realism (and therefore a possible attack on Ekert protocol).

Alice and Bob should therefore be wary of how exactly they measure the efficiency in their Quantum key distribution protocol.

If the number of emitted pair of pulses is unavailable to them, they can only estimate the conditional efficiency, by dividing the number of single counts by the number of emitted pair of pulses, and the bound is indeed 82.8$\%$. In fact, in the double blinding-attack, their estimation of the conditional efficiency would be equal to the familiar bound, that is $\eta_\mathrm{2,1}=\frac{p_\mathrm{w}}{\eta}=2(\sqrt{2}-1)\approx0.828,$ because the probability to detect a pair is simply equal to the probability $p_\mathrm{w}$ to detect a weak pulse.

If however they know the number $N$ of emitted pair of pulses and can evaluate the detection efficiency directly, by dividing the number of single counts by the number of emitted pair of pulses, then the efficiency bound is $85.3\%$.

Finally, it should be noted that if detectors with a lower efficiency are used, a simple way to deter the attack is to implement a fair sampling test \cite{Adenier08b,Adenier10FS}.

\end{document}